\newcommand{\myDocType}{Report}  
\newcommand{\myAuthor}{%
  \styleAuthor Angelo Di Porzio\\
  \styleAffiliation Scuola Superiore Meridionale
  \and
  \styleAuthor Marco Coraggio\\
  \styleAffiliation Scuola Superiore Meridionale
  }
\newcommand{\myTitle}{Reproducing Human Individual Motor Signatures: A Data-Driven Approach for Repetitive Motion}
\newcommand{\myHeaderAuthorTitle}{A.~Di Porzio et al.: Reproducing Human Individual Motor Signatures: A Data-Driven Approach for Repetitive Motion} 
\newcommand{\fixedDate}{31 July 2026}  

\PassOptionsToPackage{inline,shortlabels}{enumitem}

\documentclass[numberrefs,twocolumn,tightheaders]{lib/SMarticle}

\makeatletter
\@ifpackageloaded{natbib}{}{%
  \@ifpackageloaded{cite}{}{\RequirePackage{cite}}%
}
\makeatother

\usepackage{mathtools}
\usepackage{graphicx}
\usepackage[export]{adjustbox}

\usepackage{amsmath,amssymb,amsfonts}

\usepackage{algorithmic}
\usepackage[dvipsnames,table]{xcolor}
\usepackage{comment}
\usepackage[caption=false,font=footnotesize]{subfig}
\usepackage{bm}
\usepackage{booktabs}
\usepackage{makecell}
\usepackage{lipsum}
\usepackage{hyperref}

\providecommand{\appendices}{\appendix}

\newcommand{\sigzero}{$\text{-}$}
\newcommand{\sigone}{${*}$}
\newcommand{\sigtwo}{${*}{*}$}
\newcommand{\sigthree}{${*}{*}{*}$}

\begin{document}
\maketitle

\begin{abstract}
The deployment of autonomous virtual avatars (in extended reality) and robots in human group activities---such as rehabilitation therapy, sports, and manufacturing---is expected to increase as these technologies become more pervasive. 
Designing cognitive architectures and control strategies to drive these agents requires realistic models of human motion.
Furthermore, recent research has shown that each person exhibits a unique velocity signature, highlighting how individual motor behaviors are both rich in variability and internally consistent.
However, existing models only provide simplified descriptions of human motor behavior, hindering the development of effective cognitive architectures.
In this work, we first show that motion amplitude provides a useful characterization of individual motor signatures, complementary to existing ones.
Then, we propose a fully data-driven approach to generate original one-dimensional motion that captures the unique features of specific individuals, based on long short-term memory neural networks.
We validate the architecture using real human data from participants performing spontaneous oscillatory motion.
Thorough statistical analyses support that our model reproduces the velocity distribution and amplitude envelopes of the individual it was trained on, while remaining distinct from others.
\end{abstract}

\section{Introduction}
\label{sec:introduction}

People are naturally inclined to interact in groups during different activities, ranging from conversation to sports and craftsmanship \cite{ref:homans2017human}.
In the near future, these interactions are poised to become cyber-physical, as robots and virtual avatars join humans in shared workplaces and activities \cite{ref:cps_human_robot_colab}.
For instance, collaborative robots are expected to work alongside people in industrial processes, reducing physical and cognitive workload \cite{ref:el2017design} while enhancing social closeness within teams \cite{ref:fu2021_robot_social,ref:rennung_prosocial};
in parallel, virtual avatars in extended reality are being adopted in domains such as rehabilitation \cite{ref:howard2017meta} and sports training \cite{ref:neumann2018systematic}, guiding users through motor tasks.

To support these emerging interactions, learning-based \emph{cognitive architectures}---the advanced control strategies that enable physical and social human-robot interactions \cite{ref:langley2009cognitive}---have recently been proposed to enhance coordination in human groups through artificial agents \cite{ref:grotta} and convey affective states through motion \cite{ref:delellis_encoding}.
To synthesize, and, when machine learning is involved, train effective cognitive architectures, it is crucial to obtain accurate and realistic models of human motor dynamics and behavior \cite{ref:fang_hri_survey,ref:grotta}.

Human rhythmic motion recurs across many of these settings (sports, physiotherapy, repetitive manual tasks), making it a natural case to model.
Yet, even in these more structured tasks, characterizing and modeling human motor behavior remains challenging \cite{kelso1997dynamic,dataset_article,spallone2026rope}, because human motion is intrinsically variable: the same person never executes the same movement twice, and behavior fluctuates with attention, fatigue, and engagement.
As a methodological step toward cognitive architectures capable of supporting interaction, this work focuses on a foundational subproblem: modeling and generating motion for a single individual performing rhythmic movement in solo condition.
The model we propose is intended as a generative building block on which future architectures supporting human-machine coupling can be built.

A productive way to handle the variability of human motion is to look for the stable person-specific regularities embedded within it.
In the context of the \emph{mirror game}, pairs of participants improvising joint motion display recognizable individual patterns even while adapting to a partner \cite{ref:noy_mirror,hart2014individuality}.
Building on this evidence\cite{ref:slowinski_ims}, formalized the notion of \emph{individual motor signature} (IMS) as a stable trait characterizing human movement through its velocity profile, and showed that dynamic similarity between partners promotes interpersonal coordination.
This suggests that artificial agents capable of reproducing a target individual's IMS may provide a natural basis for cognitive architectures supporting human-machine coordination.

Earlier modeling efforts for rhythmic human motion drew primarily on coordination dynamics.
The Haken-Kelso-Bunz model \cite{haken1985theoretical} provided a foundational account of bimanual phase transitions via coupled nonlinear oscillators, and dynamical-systems approaches more broadly offer interpretable low-dimensional descriptions of rhythmic phenomena \cite{schoner1988dynamic,kelso1997dynamic}.
While powerful for theoretical analysis, these models require task-specific choices of collective variables, coupling functions, and parameters, which limits their use as personalized generative models of individual motion.
A complementary line of work is based on movement primitives, which encode motor behaviors either as attractor-based dynamical systems with learned forcing terms \cite{ijspeert2013dynamical} or as probability distributions over trajectories that support conditioning and blending \cite{paraschos2013probabilistic}.
Movement primitives are well suited to reusable, goal-directed skills learned from demonstrations, but they are not primarily designed to generate indefinite spontaneous motion that preserves individual statistical signatures.

Related to the broader goal of human-machine interaction, several works have developed artificial agents that coordinate their motion with human partners in joint motor tasks.
The human dynamic clamp paradigm \cite{kelso2009virtual,dumas2014human} couples humans with model-based virtual partners to study interpersonal coordination, typically relying on hand-designed oscillator dynamics. 
In the mirror game, \cite{zhai2016design,zhai2017design} designed virtual players by combining prerecorded human reference trajectories with feedback-control architectures, achieving effective coordination but requiring a reference motion signal. 
Along similar lines, \cite{lombardi2019deep,lombardi2021dynamic} combined reinforcement learning with motion models based on quantized representations and handcrafted multi-stage pipelines to drive cognitive architectures for joint mirror-game interaction with humans, relying on a Markovian assumption. 

In parallel, deep generative models have substantially advanced human-motion synthesis.
Recurrent neural networks, and in particular long short-term memory (LSTM) networks \cite{ref:lstm}, are well suited to modeling temporal dependencies and mitigate the vanishing-gradient issues of vanilla recurrent neural networks \cite{pmlr-v28-pascanu13}.
\cite{ref:graves_generation} showed that LSTMs combined with mixture-density outputs \cite{bishop1994mixture} can generate continuous sequences autoregressively, by sampling one step at a time from a learned conditional distribution.
Related probabilistic recurrent architectures have since been used for human trajectory prediction \cite{ref:alahi_social-lstm} and time-series generation more broadly \cite{chung2015recurrent,yoon2019timeseries}.
In computer animation, deeper generative architectures based on convolutional, normalizing-flow, and diffusion models \cite{holden2016deep,henter2020moglow,tevet2022human} have enabled full-body motion synthesis, typically conditioned on text, style categories, or control signals.
These architectures, however, are generally designed for high-dimensional pose sequences and rely on large training corpora, which does not match the realistic constraint of having only a few minutes of recording per individual. Moreover, when style is modeled, it is typically defined at the level of broad categories (e.g., ``happy walk'' or ``child-like walk'') rather than through individual idiosyncratic statistics \cite{aberman2020unpaired}, making this abstraction unsuitable for IMS reproduction from limited data.

Within this landscape, generating original, continuous, individualized rhythmic motion from a small dataset of a single individual's recordings remains an open challenge, even for simple one-dimensional motion.
We address this problem by introducing a fully data-driven architecture based on LSTM networks with Gaussian density outputs, trained on the position signals of a single person.
The model generates new motion signals autoregressively: at each step, it predicts a Gaussian distribution for the next velocity sample, samples from this distribution, filters and integrates the resulting velocity to obtain the next position value, and feeds this position back into the network, learning a continuous conditional distribution directly from raw position data.
To make autoregressive generation reliable when only limited recordings are available, we introduce a two-phase training procedure that first learns one-step-ahead velocity prediction and then selects, among the network instances produced during training, the one whose autoregressive rollouts best match the target individual's IMS.
The resulting architecture is simple, requires no handcrafted control layer or task-specific dynamical model, and performs effectively using only a few minutes of recorded data per individual.
Thorough statistical analyses confirm that each personalized model exhibits its own distinct IMS, more similar to that of the target individual than to others.

As a second contribution, we introduce motion amplitude as an additional descriptor of IMS.
We show that amplitude-based features can reveal similarities and differences in spontaneous oscillatory motion that are complementary to those captured by velocity-profile analyses \cite{ref:slowinski_ims}.
Moreover, unlike representations of velocity profiles, based on multidimensional scaling, the similarity plane of amplitudes has directly interpretable axes, and the coordinates of each signal are independent of the rest of the dataset.

The remainder of the paper is organized as follows. Section \ref{sec:preliminaries} introduces the IMS descriptors (velocity profiles and mean amplitudes). 
Section \ref{sec:problem} formalizes the problem and requirements. 
Section \ref{sec:model} details the generative architecture and training.
Section \ref{sec:results} reports the statistical validation, and Section \ref{sec:conclusions} concludes.

\section{Preliminaries}
\label{sec:preliminaries}

\begin{figure}[t]
    \centering
    \subfloat[\label{fig:signals_generation}]{%
        \includegraphics[width=\linewidth]{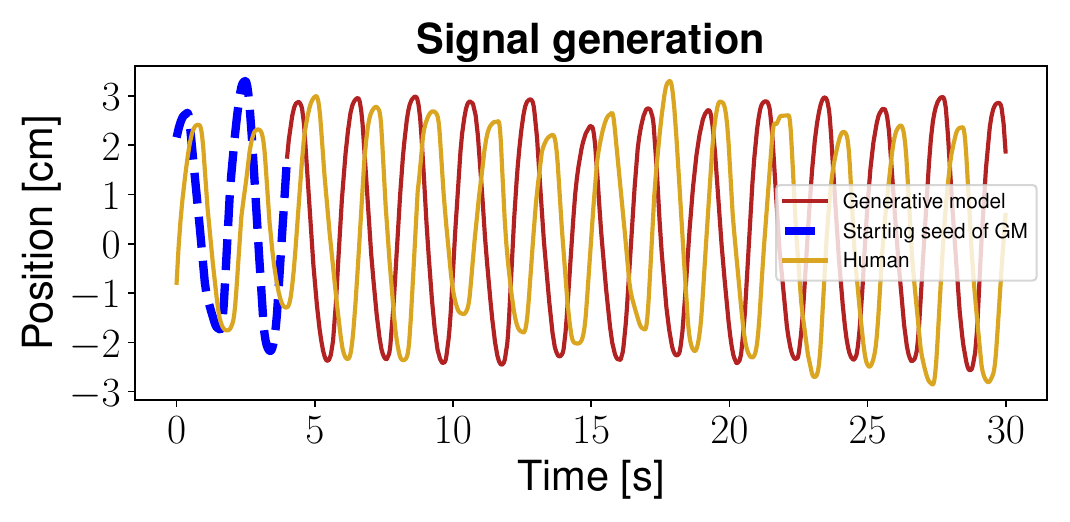}%
    } \\
    \subfloat[\label{fig:velocity_profiles_comparison}]{%
        \includegraphics[width=\linewidth]{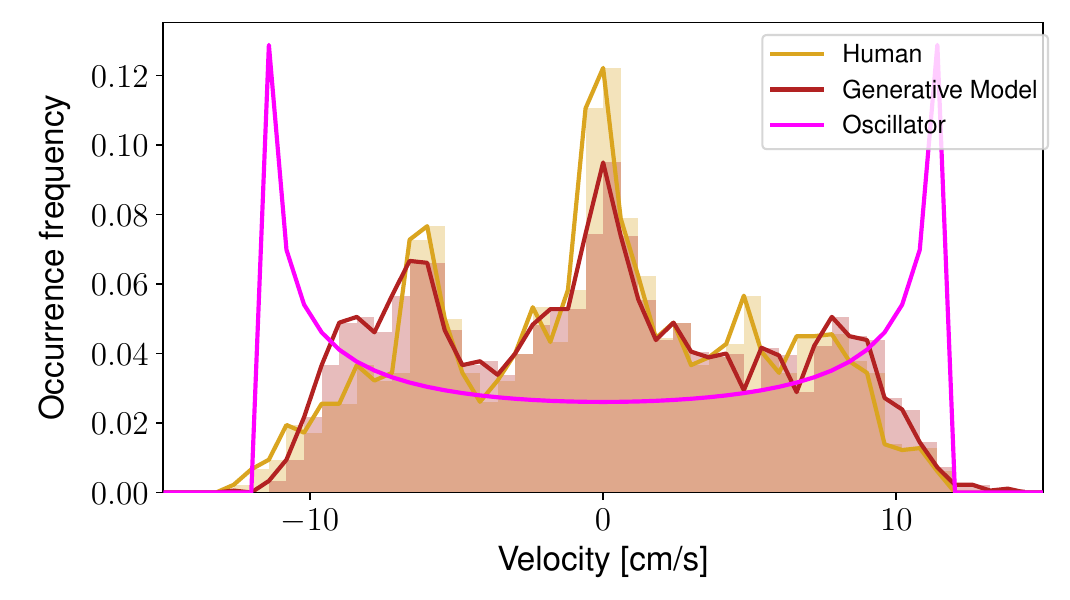}%
    }
    \caption{Comparison of (a) position timeseries and (b) velocity profiles between a human individual, the generative model (GM) trained on that individual and a linear oscillator parametrized on the individual. 
    In panel (a), the human timeseries is trial $4$ by individual $2$ (cf.~§~\ref{sec:dataset}); the timeseries from the model (cf.~§~\ref{sec:model}) is trained over individual $2$, seeded with a random $4$ s-portion of trial $4$ by individual $2$.
    Panel (b) depicts the velocity profiles of the corresponding signals, which are obtained numerically as explained in Section~\ref{sec:ims}.
    }
    \label{fig:comparison}
\end{figure}

Let $\C{T} \coloneqq \{0, 1, \dots, N_{\C{T}}-1\}$ be a finite discrete-time domain, with $N_{\C{T}} \in \BB{N}_{>0}$, and let $T_{\R{s}} \in \BB{R}_{>0}$ be the constant associated \emph{sampling time}.
In this work, we consider scalar discrete-time position signals, say $p : \C{T} \to \BB{R}$, recorded from individuals performing repetitive motor tasks, such as rhythmic limb oscillation (see, e.g., Figure \ref{fig:signals_generation}).
The velocity of a position signal $p$ is defined as 
\[
    v(t) \coloneqq \frac{p(t) - p(t-1)}{T_{\R{s}}},
\] 
setting $v(0) = 0$. 

\subsection{Individual motor signatures}
\label{sec:ims}

As shown in \cite{ref:slowinski_ims}, each person moves according to a unique \textit{individual motor signature (IMS)}; quantitatively, the IMS can be described through the velocity profiles associated to the person's movements. 
A \emph{velocity profile} is a histogram obtained by discretizing a velocity signal into bins and constructing a probability mass function over some set $\C{V}$ of uniformly distanced velocity values, with step $d_w > 0$. 
An example velocity profile of a person's oscillating motion is reported in Figure \ref{fig:velocity_profiles_comparison}; therein, for comparison, we also reported the velocity profile of a harmonic oscillator parametrized on the human motion, showing the latter fails to capture the human velocity profile (the analytical derivation is reported in Appendix \ref{appendix:harmonic_oscillator}).

The dissimilarity between two velocity profiles can be quantified using the \emph{earth mover’s distance} (EMD, corresponding to the \emph{1-Wasserstein distance}).
Namely, letting $w_1, w_2 : \C{V} \to [0, 1]$ be two velocity profiles, and letting $W_1, W_2 : \C{V} \to [0, 1]$ be their cumulative distribution functions, the EMD between $w_1$ and $w_2$ is\cite{peyré2020computationaloptimaltransport}
\begin{equation}\label{eq:emd}
    \delta^{\R{em}}\left(w_{1}, w_{2}\right) \coloneqq 
    \sum_{v \in \mathcal{V}} |W_{1}(v)-W_{2}(v)|d_w.
\end{equation}
Consistently with the notion that each person has a unique IMS, $\delta^{\R{em}}\left(w_{1}, w_{2}\right)$ tends to be lower when the velocity profiles $w_1$, $w_2$ are produced by the same person compared to the case when they are produced by different individuals.

\subsection{Multidimensional scaling}
\label{sec:multidimensional-scaling}

\begin{figure}[t]
    \centering
    \includegraphics[width=\linewidth]{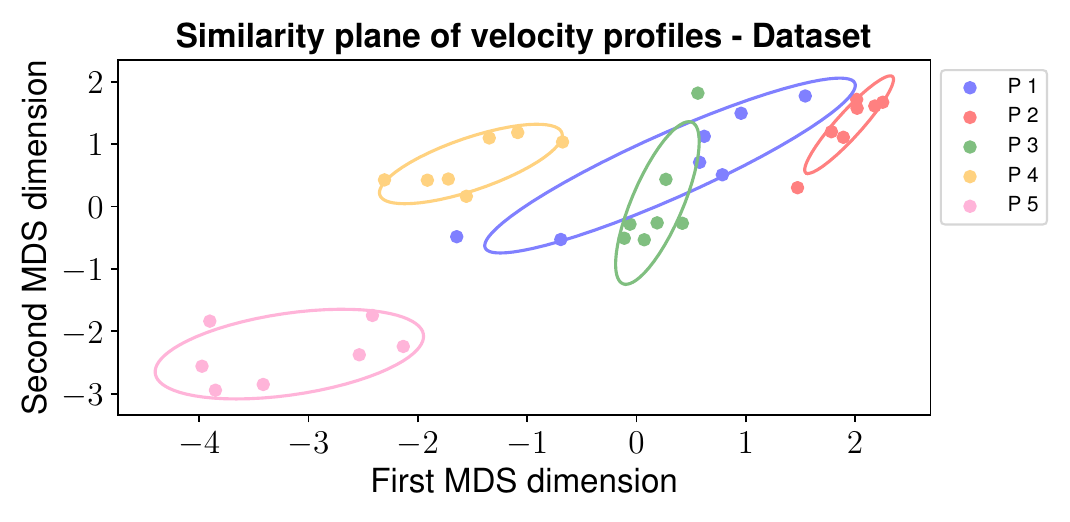} 
    \caption{Similarity plane (obtained via MDS, cf. § \ref{sec:multidimensional-scaling}) of velocity profiles for the dataset described in Section \ref{sec:dataset}. 
    Each color is associated to a human participant (P). 
    Covariance ellipses are built as explained in Appendix \ref{appendix:ellipses}.}
    \label{fig:velocity_profiles_dataset}
\end{figure}

To visualize IMSs graphically, it is possible to apply \emph{classical metric multidimensional scaling} (MDS) \cite{gower1966some,ref:borg_mds}, which  projects velocity profiles on a two-dimensional space, called the \emph{similarity plane} \cite{ref:slowinski_ims}. 
Specifically, given $n$ velocity profiles $w_1, \dots, w_n$ (even from different individuals), we let matrix $\mat{\Delta} \in \BB{R}_{\ge 0}^{n \times n}$ be given by $\Delta_{ij} = \delta^{\R{em}}\left(w_i, w_j\right)$ (e.g., see the upper left quadrant of Figure \ref{fig:EMD_velocity_profiles_dataset} below); then, applying MDS on $\mat{\Delta}$ yields a new set of points in the similarity plane ($\mathbb{R}^2$), each associated to a velocity profile, so that Euclidean distances between these points approximate the EMDs between the corresponding original velocity profiles.
An example of similarity plane for the dataset described in Section \ref{sec:dataset} is reported in Figure \ref{fig:velocity_profiles_dataset}.
Clustering of velocity profiles produced by the same person is visualized through \emph{covariance ellipses} \cite{ref:slowinski_ims}, which are built according to the procedure explained in Appendix \ref{appendix:ellipses}.

In the velocity-profile similarity plane, IMS similarity can be assessed through ellipse overlap \cite{ref:slowinski_ims} and center distances (see Appendix \ref{appendix:ellipses}). 
However, these measures are sensitive to the whole set of profiles used, since MDS depends on the full distance matrix $\mat{\Delta}$, and to the specific ellipse construction. 
Therefore, in Section \ref{sec:results}, we validate the results directly on EMDs and use similarity planes only for visualization.

\subsection{Mean amplitudes}
\label{sec:amplitudes}

As a complementary descriptor of IMS, we introduce mean amplitudes, defined as follows.

Given a position signal $p : \C{T} \to \BB{R}$ with velocity $v : \C{T} \to \BB{R}$ (cf.~Section~\ref{sec:preliminaries}), we define its \emph{positive} and \emph{negative amplitude envelopes}, denoted by $A_p^+(t), A_p^-(t)$ respectively, as
\begin{align*}
    A_p^+(t) \coloneqq 
    \begin{dcases}
        p(t), & \text{if } v(t) \le 0, v(t-1) > 0, p(t) > 0,\\
        A_p^+(t-1), & \text{otherwise},
    \end{dcases}
\\
    A_p^-(t) \coloneqq  
    \begin{dcases}
        p(t), & \text{if } v(t) \ge 0, v(t-1) < 0, p(t) < 0,\\
        A_p^-(t-1), & \text{otherwise},
    \end{dcases}
\end{align*}
setting $A_p^+(0)=0$, $A_p^-(0)=0$.
Then, we define the \emph{mean positive} and \emph{negative} \emph{amplitudes} as
\begin{equation}\label{eq:mean_amplitudes}
    \bar{A}^{+}_{p} \coloneqq \frac{1}{\lvert \mathcal{T}\rvert}\sum_{t \in  \mathcal{T}}A^{+}_{p}(t), \qquad
    \bar{A}^{-}_{p} \coloneqq \frac{1}{\lvert \mathcal{T}\rvert}\sum_{t \in  \mathcal{T}}A^{-}_{p}(t).
\end{equation}

\begin{figure}[t]
    \centering
    \includegraphics[width=\linewidth]{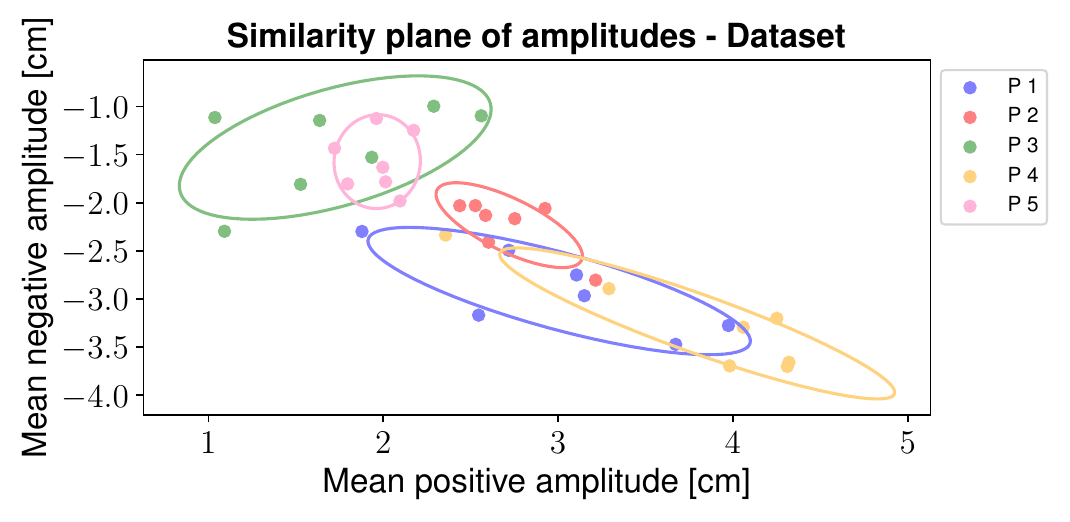} 
    \caption{Similarity plane of amplitudes (cf.~§~\ref{sec:amplitudes}) for the dataset described in Section \ref{sec:dataset}. 
    Each color is associated to a human participant (P).
    Covariance ellipses are built as explained in Appendix \ref{appendix:ellipses}.}
    \label{fig:amplitudes_dataset}
\end{figure}

Given multiple position signals, possibly from different individuals, we plot each signal by its mean positive and negative amplitudes. This defines a \emph{similarity plane of amplitudes}, analogous to the MDS-based velocity-profile similarity plane (Section \ref{sec:multidimensional-scaling}).

Figure \ref{fig:amplitudes_dataset} shows this plane for the dataset described in Section \ref{sec:dataset} below.
Covariance ellipses, computed as in Appendix \ref{appendix:ellipses}, show that signals from the same individual tend to cluster together, suggesting that motion amplitude provides a complementary descriptor of IMS.
Unlike the velocity-profile space, this plane does not rely on MDS; hence, its axes portray directly measurable physical quantities, and pairwise distances do not depend on the full set of signals considered.
The complementarity of the two descriptors is apparent by comparing Figures \ref{fig:velocity_profiles_dataset} and \ref{fig:amplitudes_dataset}.
Participants 1 and 3 overlap in the velocity-profile space but not in the amplitude space, whereas participants 1 and 4 show the opposite pattern.
Thus, velocity profiles and amplitudes capture different aspects of motion similarity.
The capability of mean amplitudes to capture subjects' distinct motor behavior is further supported by the statistical analysis in Section \ref{sec:validation_amplitudes} below (see hypothesis \ref{ite:hypothesis_well_posed_ims}).

\section{Problem Formulation}
\label{sec:problem}

We consider a dataset of scalar position signals (functions $\C{T} \to \BB{R}$) recorded from a given number of individuals performing each $n_\R{s}$ trials of repetitive motor task.
Consistently with realistic applications such as personalized physiotherapy or training---where only few minutes of recording might be available per person---we assume $n_\R{s}$ is small (i.e., a number from five to ten).

Our goal is to design the architecture of a generative model that, trained on the signals of a target person, can be rolled out autoregressively beyond the seed window to produce signals of user-specified length, satisfying the following requirements:
\begin{enumerate}[label=R\arabic*., ref=R\arabic*]
  \item\label{ite:fidelity} \emph{Fidelity.} The generated signals reproduce the target's IMS,
  being as similar to the target's own signals as the latter are to one
  another.
  \item\label{ite:specificity} \emph{Specificity.} The reproduced IMS is distinctive of the target:
  generated signals are more similar to the target's signals than to those of
  other individuals.
  \item\label{ite:originality} \emph{Originality.} The generated signals are novel rather than
  memorized; they do not collapse onto, or replay, specific recorded
  trajectories of the target.
\end{enumerate}

Because a person's IMS manifests as a \emph{distribution} of per-trial descriptors (e.g., velocity profiles, mean amplitudes), reproducing it can be cast as an \emph{empirical
distribution-matching} problem \cite{castano2022matching}.%
The objective is therefore not supervised prediction of held-out trials, but individualized generative calibration: each participant's recordings are treated as an empirical sample of that participant's IMS distribution, which the model is calibrated to reproduce.
Section \ref{sec:model} presents a fully data-driven model addressing this problem, whereas statistical formalization of the requirements (R1)--(R3) and their validation is contained in Section \ref{sec:results}.

\section{Data-driven model of human motion}
\label{sec:model}

In this section, we present the design of a data-driven machine learning model that can be trained over the position signals of an individual to generate original rhythmic motion that replicates the IMS of that person.

\subsection{Architecture of the generative model}
\label{sec:architecture}

\begin{figure*}[t]
    \centering
        \includegraphics[max width=\linewidth]{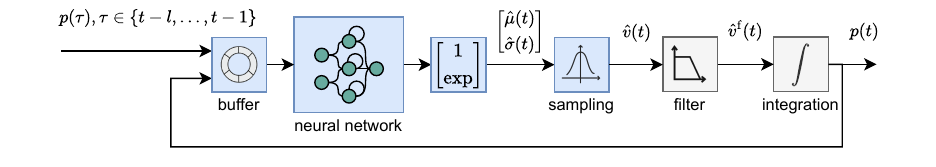}
        \caption{Schematic representation of the data-driven generative model (GM) of human motion (Section~\ref{sec:architecture}). The trained neural network receives an input position timeseries of length $l$ from a buffer and outputs $\hat{\mu}(t),\hat{\sigma}(t)$ (the exponential guarantees that $\hat{\sigma}(t) > 0$) which are interpreted as mean and standard deviation of a Gaussian distribution; the next velocity sample $\hat{v}(t)$ is sampled from that distribution; then, the velocity value is filtered and integrated to obtain a new position sample $p(t)$, which is fed back to the buffer, while the oldest sample is removed.}
    \label{fig:block_scheme}
\end{figure*}

The proposed architecture for the generative model (GM) is shown in Figure~\ref{fig:block_scheme}. 
The model takes as input $l$ initial consecutive position values $p(0), \dots, p(l-1)$ which function as seed and returns an original autoregressive sequence beyond the initial seed window, producing position values $p(t)$ for $t\ge l$.

Specifically, at time $t \geq l$, a neural network receives the last $l$ position samples, $p(t-1), \dots, p(t-l)$, and outputs two values; an exponential is applied to the latter to ensure it is positive.
These two values are the mean $\hat{\mu}(t)$ and standard deviation $\hat{\sigma}(t)$ of a Gaussian distribution from which the next velocity value $\hat{v}(t)$ is sampled.
This value is filtered using a first-order low-pass filter to 
\[
    \hat{v}^{\R{f}}(t) = (1-\beta)\hat{v}^{\R{f}}(t-1)+\beta\hat{v}(t),
\]
where $\beta \in (0, 1)$ is the filter gain.
The new position value is obtained as 
\[
    \hat{p}(t) = \hat{p}(t-1) + T_{\R{s}} \hat{v}^{\R{f}}(t).
\]
Then, $\hat{p}(t)$ is fed back into the input sequence, from which the oldest position sample $p(t-l)$ is discarded to maintain a sequence of $l$ values, which is used as new input to generate $\hat{p}(t+1)$, and so on.

Note that while in this setup position signals are assumed to be scalar, the model architecture can be used with signals of any dimension.

\subsection{Dataset}
\label{sec:dataset}

\begin{figure}[t]
    \centering
    \includegraphics[max width=\columnwidth]{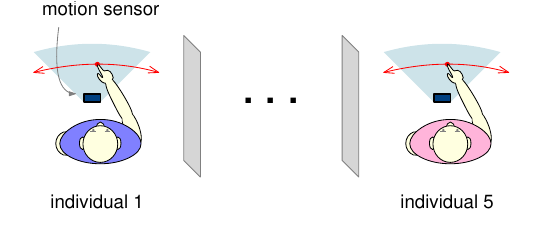}
    \caption{Schematic representation of the experimental setup considered (cf.~§~\ref{sec:dataset}).
    Five people had to move their preferred index finger continuously from left to right over a motion sensor without any interaction with other participants. 
    The motion sensor works through infrared technology and has a perception area depicted in light-blue. 
    }
    \label{fig:experiment}
\end{figure}

To train and validate our data-driven model, we use the dataset reported in \cite{dataset_article}, in which five participants were instructed to oscillate their finger continuously from left to right along a line, at a preferred frequency (see Figure~\ref{fig:experiment}).
Data were captured through Leap Motion sensors \cite{ref:guna_leapmotion} on the \textit{Chronos} platform \cite{ref:alderisio_chronos} (for more details on the experimental setup and protocol, see \cite{dataset_article}). 
The dataset consists of 7 position signals per participant, yielding a total of 35 signals.
The dataset size is consistent with previous studies on individual motor signatures and oscillatory movement \cite{ref:slowinski_ims, dataset_article} and reflects the target regime of our method: personalized applications such as physiotherapy or training, where only a few minutes of recording are available per individual; each of the five participants yields an independently trained and tested calibrated model, which constitutes the unit of validation.
Each trial lasted 30 s, was sampled at 10 Hz, interpolated via spline to 100 Hz, and processed with a Butterworth filter using a cutoff frequency twice the typical one associated with natural human movement ($\approx3$ Hz).
Consequently, the effective sampling time is $T_{\R{s}} = 0.01\,\text{s}$.
Moreover, when constructing velocity profiles (cf.~Section \ref{sec:ims}), we bin values into 101 uniform bins in the range $[-30, 30]$ cm/s, following \cite{ref:slowinski_ims}, resulting in a binning resolution of $d_w = 0.6 $ cm/s.

\subsection{Parametrization of the generative model}

As parameters for the generative model, we set the low-pass filter gain to $\beta = 0.6$ to guarantee a cut-off frequency at least 10 times the typical frequency of human natural movement ($\approx 3$ Hz \cite{dataset_article}). 
We set the length of the input window to $l = 400$ samples, which corresponds to 4 s (recalling the sampling time is $T_{\R{s}} = 0.01\,\R{s}$).
The neural network architecture comprises two LSTM layers with 20 units each, followed by one fully connected layer.
The network was trained using the Adam optimizer, with learning rate $\alpha = 1\times 10^{-4}$.
The model was implemented in Python using PyTorch \cite{ref:pytorch}.

\subsection{Training of the generative model}
\label{sec:training}

\begin{figure*}[t]
        \centering
        \includegraphics[max width=\linewidth]{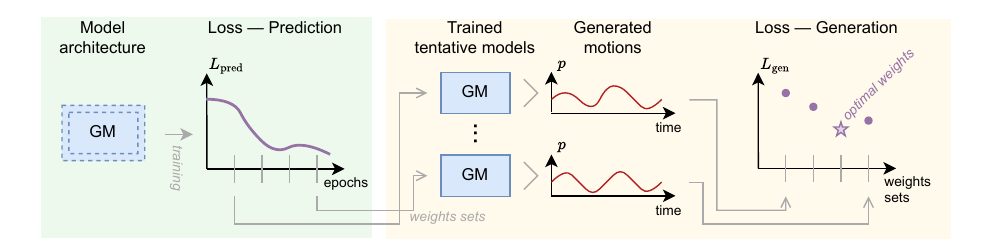}
        \caption{Schematic representation of the network weights selection process. The neural network is trained to minimize the prediction loss $L_\text{pred}$ in equation \eqref{eq:loss}; weights sets are saved during the training process and then used to generate new motion signals according to the process depicted in Figure \ref{fig:block_scheme}; the weights set that resulted in the signals which minimize the generation loss $L_\text{gen}$ (equation \eqref{eq:generation_loss}) are selected as those best representing the individual motor signature of the target person.}
        \label{fig:model_selection}
\end{figure*}

We train a separate instance of the model for each individual in the dataset, using only that participant's position signals.
The training procedure consists of two phases: 
\begin{enumerate}[label=(\roman*)]
  \item training the neural network to predict the distribution of the next velocity sample;
  \item selecting, among network instances produced during training, the one whose generated position signals best reproduce the individual's IMS.
\end{enumerate}
The two phases are described in detail below; a schematic of the training process is shown in Figure \ref{fig:model_selection}.

\subsubsection{One-step-ahead velocity prediction}

In this phase, samples consist of 4-second windows of position data $\{p(t-1), \dots, p(t-l)\}$, extracted from each human signal (of the target individual) using a sliding window over all time steps, and are used to predict the corresponding next velocity value $v(t)$.
The samples are randomly split into 70\% for training and 30\% for validation.

The neural network's parameters are learned by minimizing the negative log-likelihood of the target velocity $v(t)$ under the Gaussian distribution $\C{N}\left(\hat{\mu}(t), \hat{\sigma}(t)\right)$ \cite{ref:alahi_social-lstm}.
This yields the \emph{prediction loss} function \cite{ref:nix_nllgaussian}
\begin{equation}\label{eq:loss}
    L_{\R{pred}} = \frac{1}{2}\left(\log\left(\hat{\sigma}^{2}(t)\right)+\frac{\left(\hat{\mu}(t)-v(t)\right)^2}{\hat{\sigma}^{2}(t)}\right).
\end{equation}
Training is performed in batches of size 2000, averaging~\eqref{eq:loss} over each batch, over 8000 epochs.

\subsubsection{Autoregressive signal generation}
\label{sec:model_selection}

While minimizing~\eqref{eq:loss} optimizes one-step-ahead velocity prediction, it does not automatically guarantee that, in autoregressive generation, the model produces signals whose IMS matches that of the target individual. 
We therefore include an additional model selection stage to identify the weights that best reproduce the target IMS.

During training for the one-step-ahead velocity prediction phase, the network weights are saved every 100 epochs, yielding 80 candidate sets indexed by $h \in \{1, \dots, 80\}$. 
For each candidate set, we let the model generate $n_\R{s}$ new position timeseries, each initialized with seven 4-second pieces randomly extracted from the $n_\R{s}$ signals of the target participant. 
Let $w^{\R{d}}_{i}$ and $w^{\R{g},h}_{i}$, $i \in \{1,\dots, n_{\R{s}}\}$, denote the velocity profiles of the $n_{\R{s}} \in \mathbb{N}$ participant's signals and of those generated by the model with candidate weights set $h$, respectively.
Then, we define the \emph{generation loss} as (cf.~Section~\ref{sec:ims})
\begin{equation}\label{eq:generation_loss}
    L_\R{gen} = \frac{1}{n_{\R{s}}}\sum_{i=1}^{n_{\R{s}}}\delta^{\R{em}}\left(w^{\R{d}}_{i},w^{\R{g},h}_{i}\right).
\end{equation}
The optimal weights are selected as those minimizing $L_\R{gen}$.

\section{Results}
\label{sec:results}

In this section, we validate the model presented in Section \ref{sec:model} by assessing its ability to reproduce the motor signature of each target individual. 
Qualitative comparisons between participants' signal and generated signals are reported in Section \ref{sec:qualitative_evaluation}, while thorough statistical analyses of similarity measures of velocity profiles are contained in Section \ref{sec:statistical_analysis}.
An additional validation on the complementary IMS measure of mean amplitudes is reported in Section \ref{sec:validation_amplitudes}.

\subsection{Evaluation of the generative model}
\label{sec:qualitative_evaluation}

We evaluate the generative model by producing new position signals for each individual. 
Five models (one per participant) are trained as described in Section \ref{sec:training}.
Seven 4-second segments from each signal of each subject are extracted randomly and are used as initial seeds for each model to generate seven original signals.

Figure \ref{fig:signals_generation} compares a participant's signal (individual 2, trial 4) with a generated one.
Although original, the generated motion is consistent with the subject’s IMS, as reflected by the similar velocity profiles shown in Figure \ref{fig:velocity_profiles_comparison}.

\begin{figure}[t]
    \centering
    \includegraphics[width=\linewidth]{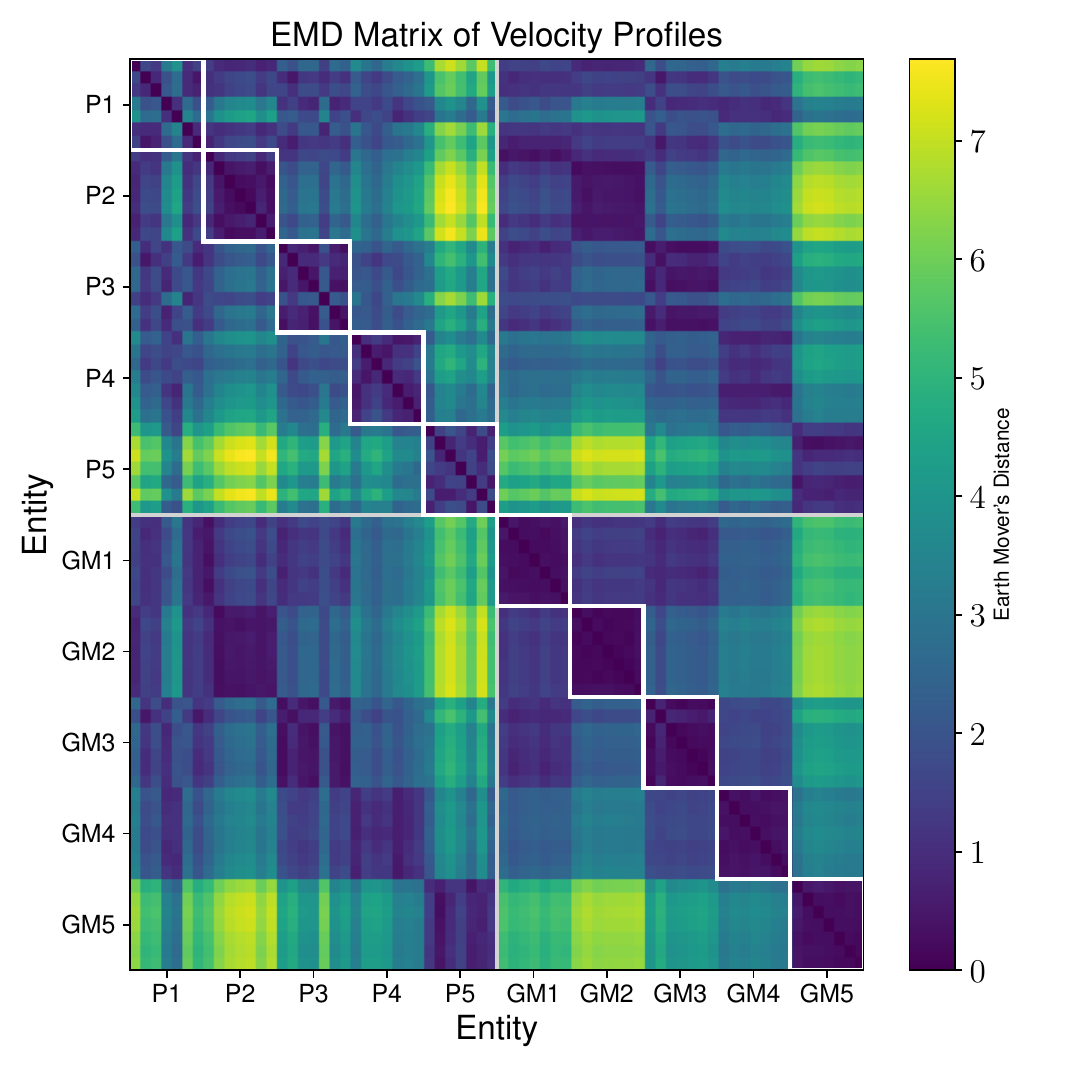} 
    \caption{Earth mover distances (EMD; cf.~\eqref{eq:emd}) matrix of signals produced by human participants (P $i$) and generative models (GM $i$). 
    Each block of $n_\R{s} \times n_\R{s}$ pixels ($n_\R{s} = 7$) collects the distances between all signal pairs from two entities (humans and/or models).
    The block-diagonal structure (white boxes) shows that signals from the same entity are more similar to one another (smaller distance) than to signals from other entities. 
    Grey lines partition the matrix into four blocks: human-human (upper left), human-models (upper right and bottom left), and models-models (bottom right).}
    \label{fig:EMD_velocity_profiles_dataset}
\end{figure}

To assess similarity at the population level, we compute the EMD between the velocity profiles of all human and generated signals, and plot the resulting EMD matrix $\mat{\Delta}$ in Figure \ref{fig:EMD_velocity_profiles_dataset}.
Each participant and each generative model is represented by $n_\R{s} = 7$ signals (i.e., 7 consecutive rows or columns), and each entry corresponds to the EMD between a pair of signals. 
The matrix exhibits a clear block-diagonal structure, with generally smaller distances within the same entity (person or model) than across different ones.
This indicates that both human and generated signals form internally consistent groups.


\begin{figure}[t]
    \centering
    \subfloat[\label{fig:similarity_space_velocity_profiles_GM}]{%
        \includegraphics[width=\linewidth]{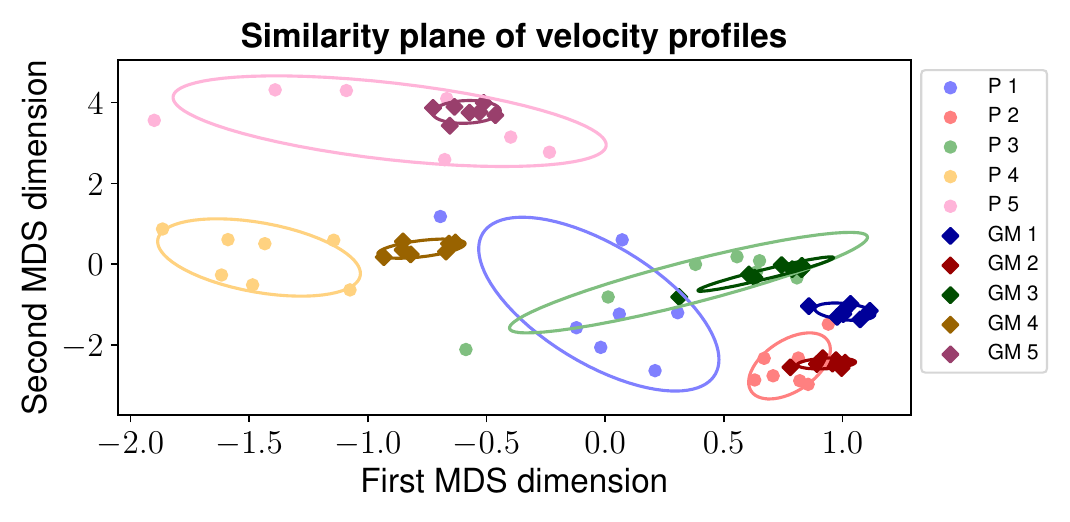}%
    } \\
    \subfloat[\label{fig:similarity_space_amplitudes_GM}]{%
        \includegraphics[width=\linewidth]{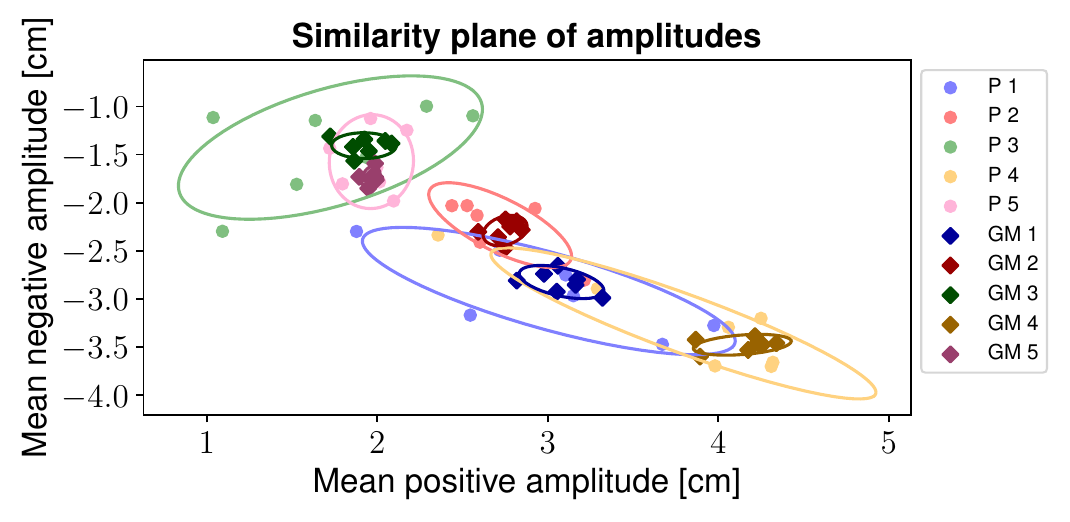}%
    }
    \caption{Similarity planes of velocity profiles (a) and mean amplitudes (b), comparing signals by human participants (P; lighter color) and generative models (GM; darker colors).}
    \label{fig:similarity_spaces_with_GMs}
\end{figure}

Figure \ref{fig:similarity_spaces_with_GMs} provides a complementary visualization by representing human and generated signals in the similarity planes.
Indeed, the covariance ellipses of the generative models are generally located near those of the corresponding participants and often overlap with them, further supporting the ability of the models to reproduce individual motor signatures.



\subsection{Statistical analysis}
\label{sec:statistical_analysis}

To further validate the results of Section \ref{sec:qualitative_evaluation}, we performed statistical tests on the EMD values shown in Figure~\ref{fig:EMD_velocity_profiles_dataset}, as described below.
This validation rests on the 35 recorded signals and their 35×35 matrix of pairwise distances. We perform 25 permutation tests—five hypotheses per participant—assessing the existence of distinct signatures, fidelity, and specificity. An additional per-model originality check is performed for each of the five models.

\subsubsection{Validation of model's fidelity and specificity}
\label{sec:validation_fidelity_specificity_emd}

Let $P_{i}$ denote the set of velocity profiles obtained from the position signals produced by participant $i$, and $G_{i}$ the set of velocity profiles obtained from the signals generated by model $i$, trained on that participant.
We also denote by $P_{\neg i}$ the set of all human velocity profiles except those of participant $i$, and by $G_{\neg i}$ the set of all generated velocity profiles except those produced by model $i$.
For any two sets of velocity profiles $A,B \in \{P_{i}, G_{i}, P_{\neg i}, G_{\neg i}\}$, we extend the notation $\delta^{\R{em}}$ (cf.~\eqref{eq:emd}) and define
\[
    \begin{aligned}
        \delta^{\R{em}}(A, B) \coloneqq & \{\delta^{\R{em}}(w_1, w_2) \mid w_1 \in A, w_2 \in B, w_1 \neq w_2 \},\\
        \delta^{\R{em},\epsilon}(A, B) \coloneqq & \{(1+\epsilon)\delta^{\R{em}}(w_1, w_2) \mid \\
        &\qquad\qquad\qquad w_1 \in A, w_2 \in B,w_1 \neq w_2 \},
    \end{aligned}
\]
that is, the set of pairwise EMDs between elements of $A$ and $B$; $\delta^{\R{em},\epsilon}$, with $\epsilon \ge 0$, is used only in hypothesis \ref{ite:hypothesis_fidelity}, which requires a tolerance margin, being the only absolute comparison. 
We then define the average of a set of distances $\delta^{\R{em}}(A,B)$ as (analogously for $\delta^{\R{em},\epsilon}_{\R{avg}}(A,B)$)
\begin{equation}
\label{eq:avergage_emd_sets}
    \delta^{\R{em}}_{\R{avg}}(A, B) \coloneqq  \frac{1}{\lvert \delta^{\mathrm{em}}(A, B) \rvert} \sum_{\xi \in \delta^{\R{em}}(A, B)} \xi.
\end{equation}

Using $\delta^{\R{em}}_{\R{avg}}$ as summary statistics, we tested the following hypotheses:
\begin{enumerate}[label=H\arabic*., ref=H\arabic*]
    \item\label{ite:hypothesis_well_posed_ims} Subjects' IMSs are distinct; supported by the fact that signals from participant $i$ are more similar to one another than to those of other participants, i.e.,
    \[
    \delta^{\mathrm{em}}_{\mathrm{avg}}(P_i,P_i) \le \delta^{\mathrm{em}}_{\mathrm{avg}}(P_i,P_{\neg i}).
    \]
    
    \item\label{ite:hypothesis_fidelity} A model reproduces the IMS of its associated individual; supported by the fact that signals generated by model $i$ are at least as similar to those of participant $i$ as the latter are to one another,%
    \footnote{Unlike the other hypotheses, which compare which of two distances is smaller, \ref{ite:hypothesis_fidelity} requires the generated-to-target distance to meet an absolute reference---the target's own intra-individual distance---so we admit a non-zero tolerance $\epsilon$.}
    i.e.,
    \[ 
        \delta^{\mathrm{em}}_{\mathrm{avg}}(P_i,G_i) \le \delta^{\mathrm{em},\epsilon}_{\mathrm{avg}}(P_i,P_i).
    \]

    \item\label{ite:hypothesis_models_are_distinct} Each model exhibits a distinct signature; supported by the fact that signals generated by model $i$ are more similar to one another than to those generated by other models, i.e.,
    \[
    \delta^{\mathrm{em}}_{\mathrm{avg}}(G_i,G_i) \le \delta^{\mathrm{em}}_{\mathrm{avg}}(G_i,G_{\neg i}).
    \]
    
    \item\label{ite:hypothesis_specificity_person_to_model} An individual's IMS is best captured by their corresponding model (rather than other models); supported by the fact that signals from participant $i$ are more similar to those generated by the corresponding model than to those generated by other models, i.e.,
    \[
    \delta^{\mathrm{em}}_{\mathrm{avg}}(P_i,G_i) \le \delta^{\mathrm{em}}_{\mathrm{avg}}(P_i,G_{\neg i}).
    \]
    
    \item\label{ite:hypothesis_specificity_model_to_person} A model specifically reproduces the target participant's IMS (rather than generic human motion); supported by the fact that signals generated by model $i$ are more similar to those of the corresponding participant than to those of other participants, i.e.,
    \[
        \delta^{\mathrm{em}}_{\mathrm{avg}}(P_i,G_i) \le \delta^{\mathrm{em}}_{\mathrm{avg}}(P_{\neg i},G_i).
    \]   
\end{enumerate}

Hypothesis \ref{ite:hypothesis_well_posed_ims} confirms that the problem is well posed, i.e., that distinct IMSs exist in the data to be reproduced;
\ref{ite:hypothesis_fidelity} validates \emph{fidelity} (requirement~\ref{ite:fidelity});
\ref{ite:hypothesis_models_are_distinct} establishes that the models are mutually distinct and is precondition for validating \emph{specificity} (\ref{ite:specificity}), which is validated by \ref{ite:hypothesis_specificity_person_to_model} and \ref{ite:hypothesis_specificity_model_to_person}.
\emph{Originality} (\ref{ite:originality}) is assessed separately in Section \ref{sec:originality_testing}.

Because model selection in Section \ref{sec:model_selection} chooses weights minimizing the generation loss against the target's own profiles, \ref{ite:hypothesis_fidelity} is a \emph{calibration} check rather than held-out prediction: it verifies the selected model attains the target's intra-individual distance.
The discriminative claims---distinctness (\ref{ite:hypothesis_models_are_distinct}), correct attribution across models and participants (\ref{ite:hypothesis_specificity_person_to_model}, \ref{ite:hypothesis_specificity_model_to_person}), and non-memorization (Section~\ref{sec:originality_testing})---are not directly optimized by the selection criterion and therefore carry the specificity evidence.

For Hypothesis \ref{ite:hypothesis_fidelity}, we set the margin $\epsilon$ to the sampling uncertainty of the reference average: under a faithful model, the mean generated-to-target distance and the mean intra-individual distance coincide in expectation, so the two observed averages differ only through finite-sample fluctuation.
Therefore, testing the $i$-th model, we take $\epsilon = \frac{\R{SE}_i}{\delta_{\R{avg}}^{\R{em}}(P_i,P_i)}$, where $\R{SE}_i$ is the standard error of $\delta_{\R{avg}}^{\R{em}}(P_i,P_i)$. 
Since the pairwise distances in ${\delta_{\R{avg}}^{\R{em}}(P_i,P_i)}$ are not independent, we estimate $\R{SE}_i$ by a jackknife over signals \cite{efron1981nonparametric, efron1981jackknife}.%
\footnote{Leaving out one signal $k$ at a time, we let $\bar\delta_{(-k)} = \binom{n_s-1}{2}^{-1}\!\sum_{l<m,\ l,m\neq k} \delta^{\R{em}}(w_l,w_m)$, with $w_l,w_m \in P_i$, and we let $\bar\delta_{(\cdot)} = \frac{1}{n_s}\sum_{k=1}^{n_s}\bar\delta_{(-k)}$.
Then, we compute $\R{SE}_i = \big[\frac{n_s-1}{n_s}\sum_{k=1}^{n_s}(\bar\delta_{(-k)}-\bar\delta_{(\cdot)})^2\big]^{1/2}$.}

To perform the statistical tests, we used permutation testing.
Under the null hypothesis of exchangeability, signals were randomly reassigned between the two groups, and the test statistic ($\delta^{\R{em}}_{\R{avg}}$) was recomputed for each permutation to construct an empirical null distribution.
This resampling procedure was repeated 5000 times. 
The $p$-value was estimated from the empirical null distribution according to the specified alternative hypothesis, as the proportion of permuted statistics less than or equal to the observed statistic. 
Each hypothesis was tested separately for each participant, yielding a total of 25 tests.
The resulting $p$-values were corrected for multiple comparisons using the Benjamini-Yekutieli procedure, in order to account for dependencies among tests \cite{benjamini2001control}.

\begin{table}[t]
\centering
\begin{tabular}{l|ccccc}
\toprule
\textbf{Person} & \textbf{H1} & \textbf{H2} & \textbf{H3} & \textbf{H4} & \textbf{H5} \\
\midrule
P1 &
\makecell[c]{\sigone \\ 0.0247} &
\makecell[c]{\sigtwo \\  0.0023} &
\makecell[c]{\sigthree \\  0.0005} &
\makecell[c]{\sigthree \\  0.0005} &
\makecell[c]{\sigthree \\  0.0005} \\
\arrayrulecolor{gray!40}\midrule
P2 &
\makecell[c]{\sigthree \\  0.0006} &
\makecell[c]{\sigtwo \\  0.0015} &
\makecell[c]{\sigthree \\  0.0005} &
\makecell[c]{\sigthree \\  0.0005} &
\makecell[c]{\sigthree \\  0.0005} \\
\midrule
P3 &
\makecell[c]{\sigthree \\  0.0006} &
\makecell[c]{\sigtwo \\  0.0015} &
\makecell[c]{\sigthree \\  0.0005} &
\makecell[c]{\sigthree \\  0.0005} &
\makecell[c]{\sigthree \\  0.0005} \\
\midrule
P4 &
\makecell[c]{\sigthree \\  0.0006} &
\makecell[c]{\sigzero \\  0.1543} &
\makecell[c]{\sigthree \\  0.0005} &
\makecell[c]{\sigthree \\  0.0005} &
\makecell[c]{\sigthree \\  0.0005} \\
\midrule
P5 &
\makecell[c]{\sigthree \\  0.0006} &
\makecell[c]{\sigtwo \\  0.0015} &
\makecell[c]{\sigthree \\  0.0005} &
\makecell[c]{\sigthree \\  0.0005} &
\makecell[c]{\sigthree \\  0.0005} \\
\arrayrulecolor{black}\bottomrule
\end{tabular}

\caption{$p$-values of statistical permutation tests applied on EMD (cf.~§~\ref{sec:validation_fidelity_specificity_emd}).
$p$-values are corrected via a false discovery control with a Benjamini-Yekutieli procedure. Significance is reported as \sigone \ for $p<0.05$, \sigtwo \ for $p<0.01$, and \sigthree \ for $p<0.001$. No significance ($p\geq 0.05$) is reported as \sigzero.}
\label{tab:p-values-emd}
\end{table}

The resulting $p$-values are reported in Table~\ref{tab:p-values-emd}.
Hypotheses \ref{ite:hypothesis_well_posed_ims}, \ref{ite:hypothesis_models_are_distinct}, \ref{ite:hypothesis_specificity_person_to_model}, and \ref{ite:hypothesis_specificity_model_to_person}  are supported for all individuals. 
These results support the existence of individual motor signatures in the human data (\ref{ite:hypothesis_well_posed_ims}), show that the generative models also exhibit distinct signatures (\ref{ite:hypothesis_models_are_distinct}), and indicate that each model produces a signature that is more similar to the corresponding target individual than those of the other models (\ref{ite:hypothesis_specificity_person_to_model}); conversely, the target individual’s signature is the most similar, among those in the dataset, to that of the corresponding model (\ref{ite:hypothesis_specificity_model_to_person} ). 
Hypothesis \ref{ite:hypothesis_fidelity} holds for 4 out of 5 participants within the one-standard-error tolerance ($\epsilon$), confirming that the model is in most cases able to capture the target's IMS.
For completeness, we report that hypothesis \ref{ite:hypothesis_fidelity} holds also for participant P4 when computed with a tolerance of $1.5$ (rather than one) standard errors ($p = 0.0310$).


\subsubsection{Validation of model's originality}
\label{sec:originality_testing}

To verify that the model generates novel signals rather than simply replaying recorded ones, we check that the generated signals are not pathologically close to those of the target individual, as follows.

For each individual $i$ and each generated velocity profile of its model, $w_j^{G_i} \in G_i$, let $w_\R{near}^{P_i}(w_j^{G_i}) \in P_i$ be the nearest signal from the target individual, and $\gamma^{G\text{-}P}(w_j^{G_i})$ be the distance between the two, i.e.,
\begin{align*}
    w_\R{near}^{P_i}(w_j^{G_i}) &\coloneqq \arg\min_{w \in P_i} \delta^{\mathrm{em}}(w_j^{G_i}, w),\\
    \gamma^{G\text{-}P}(w_j^{G_i}) &\coloneqq \phantom{\arg}\min_{w \in P_i} \delta^{\mathrm{em}}(w_j^{G_i}, w),
\end{align*}
Let $\gamma^{P\text{-}P}(w_j^{G_i})$ be the distance from $w_\R{near}^{P_i}(w_j^{G_i})$ to the nearest signal of the same $i$-th individual, i.e.,
\begin{equation*}
    \gamma^{P\text{-}P}(w_j^{G_i}) \coloneqq \min_{w \in P_i, w \neq w_\R{near}^{P_i}(w_j^{G_i})} \delta^{\mathrm{em}}(w_\R{near}^{P_i}(w_j^{G_i}), w).
\end{equation*}
We then average the ratio of these distances over all signals generated by the $i$-th model, obtaining
\begin{equation*}
    \rho^{G_i} \coloneqq \frac{1}{n_{\R{s}}}\sum_{j=1}^{n_{\R{s}}} \frac{\gamma^{G\text{-}P}(w_j^{G_i})}{\gamma^{P\text{-}P}(w_j^{G_i})}.
\end{equation*}

A value $\rho^{G_i} \approx 1$ means that, on average, signals produced by the $i$-th model sit as far from the nearest recording of the target individual as such recordings sit from other recordings of the individual; conversely, $\rho^{G_i} \ll 1$ (i.e., close to $0$) would indicate that generated signals collapse onto specific individual's recordings.
Originality therefore requires $\rho^{G_i}$ to remain of order $1$ or above.

The computed ratios for the trained models are reported in Table~\ref{tab:ratios-emd}, which shows all ratios above $0.75$, confirming that no model pathologically replicates a specific subset of its target's recordings; hence requirement \ref{ite:originality} is satisfied for all individuals.

\begin{table}[t]
\centering
\setlength{\tabcolsep}{5pt}
\renewcommand{\arraystretch}{1.2}

\begin{tabular}{l|ccccc}
\toprule
 & \textbf{P1} & \textbf{P2} & \textbf{P3} & \textbf{P4} & \textbf{P5} \\
\midrule
$\rho^{G_i}$ & 0.7575 & 0.8801 & 0.9115 & 1.0624 & 0.8000 \\
\bottomrule
\end{tabular}

\caption{Results of the originality test detailed in § \ref{sec:originality_testing}.
$\rho^{G_i} \approx 1$ indicates the $i$-th model generate signals as original as the target individual; $\rho^{G_i} \ll 1$ (i.e., close to $0$) would indicate pathological collapse of the model on the dataset.}
\label{tab:ratios-emd}
\end{table}

\subsection{Further validation on amplitude envelopes}
\label{sec:validation_amplitudes}

In Section~\ref{sec:amplitudes}, we introduced mean amplitudes as a descriptor of IMS complementary to velocity profiles.
Here, we verify (i) that this descriptor separates individuals---and is therefore a valid IMS measure in its own right---and (ii) that the generative model reproduces each target's IMS in the amplitude space as well.

To quantify the dissimilarity between two position signals $p_1$ and $p_2$ in terms of amplitude, we define
\begin{equation}\label{eq:mean_amplitude_distance}
    \delta^{\R{a}}\left(p_{1}, p_{2}\right) \coloneqq
    \sqrt{\left(\bar{A}^{+}_{p_1} - \bar{A}^{+}_{p_2}\right)^2 + \left(\bar{A}^{-}_{p_1} - \bar{A}^{-}_{p_2}\right)^2},
\end{equation}
where $\bar{A}^{+}_{p}$ and $\bar{A}^{-}_{p}$ are the mean amplitudes defined in Section~\ref{sec:amplitudes}.

\begin{figure}[t]
    \centering
    \includegraphics[width=\linewidth]{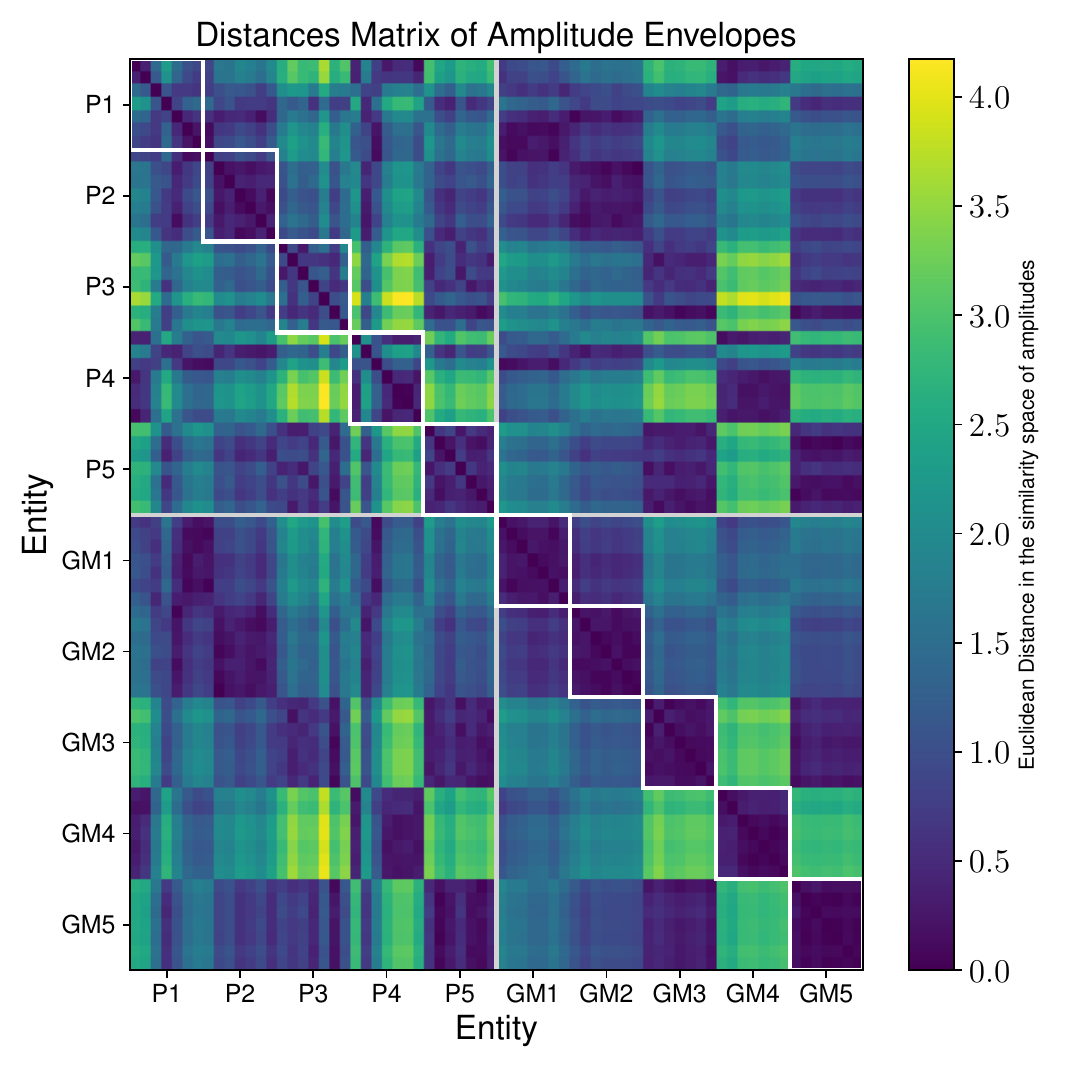} 
    \caption{Matrix of mean amplitude distances (cf.~\eqref{eq:mean_amplitude_distance}) for signals produced by human participants (P $i$) and generative models (GM $i$). 
    Each block of $n_\R{s} \times n_\R{s}$ pixels ($n_\R{s} = 7$) collects the distances between all signal pairs from two entities (humans and/or models).
    The block-diagonal structure (white boxes) shows that signals from the same entity are more similar to one another (smaller distance) than to signals from other entities. 
    Grey lines partition the matrix into four blocks: human-human (upper left), human-models (upper right and bottom left), and models-models (bottom right).}
    \label{fig:amplitude_distance_matrix}
\end{figure}

The Euclidean distance matrix in the amplitude similarity plane, shown in Figure \ref{fig:amplitude_distance_matrix}, mirrors the EMD matrix (Figure \ref{fig:EMD_velocity_profiles_dataset}): each pixel reports the distance $\delta^\R{a}$
 between the two corresponding signals.
 A block-diagonal structure is again visible, showing that signals produced by the same person stay reciprocally closer than any others.

To further validate the data-driven model of human motion  proposed in Section \ref{sec:model}, we repeat the analysis from Section \ref{sec:statistical_analysis} with $\delta^\R{a}$ in place of $\delta^{\R{em}}$.
Following Section \ref{sec:statistical_analysis}, we extend $\delta^\R{a}$ to the set of pairwise amplitude distances between the position signals in any two sets $A$, $B$ as
\[
\begin{aligned}
    \delta^{\R{a}}(A, B) &\coloneqq \{\delta^{\R{a}}(p_1, p_2) \mid p_1 \in A, p_2 \in B, p_1 \ne p_2 \},\\
    \delta^{\R{a},\epsilon}(A, B) &\coloneqq \{(1+\epsilon)\delta^{\R{a}}(p_1, p_2) \mid p_1 \in A, p_2 \in B, p_1 \ne p_2 \},
\end{aligned}
\]
and define the average of $\delta^{\R{a}}(A, B)$ (analogously for $\delta^{\R{a},\epsilon}_{\R{avg}}$)
\begin{equation*}
    \delta^{\R{a}}_{\R{avg}}(A, B) \coloneqq  \frac{1}{\lvert \delta^{\mathrm{a}}(A, B) \rvert} \sum_{\xi \in \delta^{\R{a}}(A, B)} \xi.
\end{equation*}
We then run the same tests performed in Section~\ref{sec:statistical_analysis}, using $\delta^{\R{a}}_{\R{avg}}$ and $\delta^{\R{a},\epsilon}_{\R{avg}}$ as the summary statistics in place of $\delta^{\R{em}}_{\R{avg}}$ and $\delta^{\R{em},\epsilon}_{\R{avg}}$, with the sets $P_{i}, G_{i}, P_{\neg i}, G_{\neg i}$ now containing position signals rather than their velocity profiles.

\begin{table}[t]
\centering
\setlength{\tabcolsep}{5pt}
\renewcommand{\arraystretch}{1.2}

\begin{tabular}{l|ccccc}
\toprule
\textbf{Person} & \textbf{H1} & \textbf{H2} & \textbf{H3} & \textbf{H4} & \textbf{H5} \\
\midrule
P1 &
\makecell[c]{\sigtwo \\ 0.0046} &
\makecell[c]{\sigthree \\ 0.0009} &
\makecell[c]{\sigthree \\ 0.0005} &
\makecell[c]{\sigthree \\ 0.0005} &
\makecell[c]{\sigthree \\ 0.0005} \\
\arrayrulecolor{gray!40}\midrule
P2 &
\makecell[c]{\sigthree \\ 0.0008} &
\makecell[c]{\sigthree \\ 0.0009} &
\makecell[c]{\sigthree \\ 0.0005} &
\makecell[c]{\sigthree \\0.0005} &
\makecell[c]{\sigthree \\ 0.0005} \\
\midrule
P3 &
\makecell[c]{\sigtwo \\ 0.0011} &
\makecell[c]{\sigthree \\ 0.0009} &
\makecell[c]{\sigthree \\ 0.0005} &
\makecell[c]{\sigthree \\ 0.0005} &
\makecell[c]{\sigthree \\ 0.0005} \\
\midrule
P4 &
\makecell[c]{\sigthree \\ 0.0008} &
\makecell[c]{\sigthree \\ 0.0009} &
\makecell[c]{\sigthree \\ 0.0005} &
\makecell[c]{\sigthree \\ 0.0005} &
\makecell[c]{\sigthree \\ 0.0005} \\
\midrule
P5 &
\makecell[c]{\sigthree \\ 0.0008} &
\makecell[c]{\sigthree \\ 0.0009} &
\makecell[c]{\sigthree \\ 0.0005} &
\makecell[c]{\sigthree \\ 0.0005} &
\makecell[c]{\sigthree \\ 0.0005} \\
\arrayrulecolor{black}\bottomrule
\end{tabular}

\caption{$p$-values of statistical permutation tests applied on distances in mean amplitude (cf.~§~\ref{sec:validation_amplitudes}). 
$p$-values are corrected via a false discovery control with a Benjamini-Yekutieli procedure.
Significance is reported as \sigone\ for $p<0.05$, \sigtwo\ for $p<0.01$, and \sigthree\ for $p<0.001$.}
\label{tab:p-values-amp}
\end{table}

The resulting $p$-values are summarized in Table~\ref{tab:p-values-amp}; following the statistical analysis described in Section \ref{sec:statistical_analysis}, the tolerance $\epsilon$ used to test hypothesis \ref{ite:hypothesis_fidelity} for the $i$-th model is selected as in Section \ref{sec:validation_fidelity_specificity_emd} by taking $\epsilon = \frac{\R{SE}_i}{\delta_{\R{avg}}^{\R{a}}(P_i,P_i)}$, where $\R{SE}_i$ is the standard error of $\delta_{\R{avg}}^{\R{a}}(P_i,P_i)$ and is estimated by a jackknife over signals.
Hypothesis \ref{ite:hypothesis_well_posed_ims} holds for all individuals, confirming that amplitudes separate the human data and thus serve as a complementary measure of IMS.
Hypotheses \ref{ite:hypothesis_fidelity}--\ref{ite:hypothesis_specificity_model_to_person}  are likewise confirmed:
the model reproduces each target's IMS in amplitude terms (\ref{ite:hypothesis_fidelity}; requirement \ref{ite:fidelity}: fidelity), the models remain mutually distinct (\ref{ite:hypothesis_models_are_distinct}), and each model's signature matches its own target specifically (\ref{ite:hypothesis_specificity_person_to_model} and \ref{ite:hypothesis_specificity_model_to_person}: requirement \ref{ite:specificity}, specificity).%
\footnote{Originality (requirement \ref{ite:originality}) is not tested in terms of amplitude, as even a perfect amplitude match between a generated and a recorded signal would only indicate that the two share the same mean amplitudes, not that the generated signal replicates exactly the recorded one.}



\section{Conclusions}
\label{sec:conclusions}

Modeling human motion behavior is essential for designing robots and virtual-reality avatars that collaborate with people in motor tasks.
In this work, we addressed this problem for spontaneous rhythmic movement. 
Even in one dimension, the task is non-trivial, as human motion is inherently variable and a person  never performs the same movement twice; an effective model must therefore reproduce the personal variability as encapsulated in the individual motor signature (IMS).

We proposed a data-driven model that predicts the distributions of future velocity samples from position timeseries. 
In addition, we introduced amplitude envelopes as a new descriptor of IMS that complements existing velocity profiles description.
Statistical tests support target-specific IMS reproduction for an oscillatory motor task: model specificity is confirmed for all participants in both velocity-profile and amplitude spaces, originality tests indicate no collapse onto recorded trajectories, while fidelity is fully supported in amplitude space and strongly supported in velocity-profile space, with only one model instance not satisfying the strictest one-standard-error tolerance.
Our validation is confined to five participants performing a one-dimensional oscillatory task; this suffices to demonstrate per-individual calibration but leaves validation across larger populations and richer motor tasks open for future work.

The main direction for prospective research is to extend the model to respond to external stimuli, which is essential for group settings, where individuals continuously adapt their motion to others.
Such interaction would let networks of generative agents emulate human collective behavior, such as synchronized movement or joint action, providing a basis for the training of the machine-learning-based cognitive architectures that drive collaborative avatars and robots.
A second direction is applying the architecture to higher-dimensional repetitive movements, such as complex rehabilitation and sports exercises.
Indeed, the proposed model imposes no constraints on the dimensionality of the position data, though this step will require  IMS metrics suited to higher-dimensional motion, which are still not established in the literature.

\appendices

\section{Covariance ellipses and similarity metrics}
\label{appendix:ellipses}

To geometrically identify clusters of points corresponding to an individual in the similarity plane, covariance ellipses are used. These are derived from bivariate Gaussian distributions fitted to that person’s data points in the similarity plane  \cite{ref:slowinski_ims}.
The center $c_k$ of the ellipse for participant $k$ is the mean of their data points.
The directions of the ellipse's semi-axes are given by the eigenvectors of the covariance matrix of the data points, while their lengths are determined as the product of the square root of the eigenvalues times the Mahalanobis radius.
Namely, let \( H \) be the cumulative distribution function of a chi-squared distribution with 2 degrees of freedom.
The Mahalanobis radius we select is defined as \( R = \sqrt{\chi^2_{0.7}} \), where \( \chi^2_{0.7} = H^{-1}(0.7) \), in order to enclose 70\% of the probability mass of a bivariate Gaussian distribution. 
Smaller ellipses indicate more consistent motor behavior, while larger ellipses suggest greater within-subject variability (see, e.g., Figure~\ref{fig:velocity_profiles_dataset}).

In \cite{ref:slowinski_ims}, overlap has been proposed to quantify the similarity between individual motor signatures. 
Namely, letting $E_{i} \subset \BB{R}^2$ denote the ellipse associated to individual $i$, the \emph{overlap} $\Omega_{ij} \in [0, 1]$ between the ellipses of individuals $i$ and $j$ is given by
\begin{equation*}
    \Omega_{ij} \coloneqq \frac{m_L(E_{i} \cap E_{j}) }{ m_L(E_{i} \cup E_{j})},
\end{equation*}
where $m_L(\cdot)$ is the Lebesgue measure.
A small $\Omega_{ij}$ indicates distinct motor behaviors, whereas a large $\Omega_{ij}$ suggests greater similarity.
Because small, nearby ellipses may have zero overlap despite similar IMSs, the \emph{center distance} can provide a complementary similarity measure:
\begin{equation*}
    \delta_{ij} \coloneqq \norm{c_i-c_j}_2,
\end{equation*}
where $c_i$ and $c_j$ are the ellipse centers. 
The smaller $\delta_{ij}$ is, the more similar the IMSs of participants $i$ and $j$ are.


\section{Velocity profile of a harmonic oscillator}
\label{appendix:harmonic_oscillator}

We derived an analytical expression of the velocity profile for a harmonic oscillator, shown in Figure~\ref{fig:velocity_profiles_comparison}, using the transformation law of random variables \cite{wasserman2004all}.

Consider the oscillator's position
\[
p(t)=A\sin(\omega t),
\]
with amplitude $A>0$ and angular frequency $\omega>0$, parametrized to match the average amplitude and frequency of the human position signal in Figure~\ref{fig:velocity_profiles_comparison}.
Its velocity is
\[
    v(t)=\frac{\mathrm{d}p}{\mathrm{d}t}(t)=A\omega\cos(\omega t).
\]
Let $T$ and $V$ be the random variables associated with time and velocity, with realizations $t$ and $v$, respectively. Denote the velocity transformation by $g : t \mapsto v$, with inverse $g^{-1} : v \mapsto t$, when it exists.
We first derive the velocity profile (i.e., the probability density function $f_V$) over an interval where $g(t) = v(t)$ is monotone, so that the transformation law can be applied.
Specifically, restrict time to $[0,\pi/\omega]$, with density
\begin{equation*}
    f_T(t) = \begin{cases}
        \frac{\omega}{\pi}, \quad &t\in [0, \frac{\pi}{\omega}], \\
        0, \quad &\text{elsewhere}.
    \end{cases}
\end{equation*}
Then, $V$ is obtained by sampling $v(t)$ uniformly in time, and the inverse velocity law is, for $v \in [-A\omega, A\omega]$,
\[
    t = g^{-1}(v) = \frac{1}{\omega} \arccos \left(\frac{v}{A \omega}\right).
\]
Applying the transformation of random variables yields, for $v \in [-A\omega, A\omega]$,
\begin{equation}
    \begin{aligned}
\label{eq:pdf_oscillator}
    f_V(v) &= 
    f_T\left(g^{-1}(v)\right) \left\lvert \frac{\R{d}}{\R{d}v}g^{-1}(v)\right\rvert\\
    &=
    \frac{\omega}{\pi}\left\lvert \frac{1}{\omega}\frac{\text{d}}{\text{d}v}\arccos \left(\frac{v}{A\omega}\right) \right\rvert
    =
    \frac{1}{\pi A\omega }\frac{1}{\sqrt{1-\left(\frac{v}{A\omega}\right)^2}}.
\end{aligned}
\end{equation}

By symmetry, the density derived on the monotone branch $[0, \frac{\pi}{\omega}]$ extends to one full period $[0, \frac{2\pi}{\omega}]$: indeed, the same velocity values recur over $[\frac{\pi}{\omega}, \frac{2\pi}{\omega}]$ with the same relative time frequency. 
By periodicity, repeating the argument over multiple periods does not change these relative frequencies. 
Therefore, the analytical form in \eqref{eq:pdf_oscillator} describes the oscillator velocity profile over arbitrary full-period observation windows.

\bibliographystyle{IEEEtran} 
\bibliography{bibliography}

\end{document}